\documentclass[aip,
 amsmath,amssymb,
 reprint,%
]{revtex4-1}

\usepackage{graphicx}
\usepackage{dcolumn}
\usepackage{bm}

\usepackage[utf8]{inputenc}
\usepackage[T1]{fontenc}
\usepackage{mathptmx}
\usepackage{etoolbox}

\makeatletter
\def\@email#1#2{%
 \endgroup
 \patchcmd{\titleblock@produce}
  {\frontmatter@RRAPformat}
  {\frontmatter@RRAPformat{\produce@RRAP{*#1\href{mailto:#2}{#2}}}\frontmatter@RRAPformat}
  {}{}
}%
\makeatother
\begin{document}

\title{Surface viscosity in simple liquids}

\author{Paolo Malgaretti}
\affiliation{Forschungszentrum Jülich GmbH, Helmholtz Institute Erlangen-Nürnberg for Renewable Energy (IEK-11), Cauerstr.1, D-91058 Erlangen, Germany}
\author{Ubaldo Bafile}
\author{Renzo Vallauri}
\affiliation{Consiglio Nazionale delle Ricerche, Istituto di Fisica Applicata ``Nello Carrara'', I-50019 Sesto Fiorentino, Italy}
\author{P\'al Jedlovszky}
\affiliation{Department of Chemistry, Eszterházy Károly University, Leányka u. 6, H-3300 Eger, Hungary}
\author{Marcello Sega}
\affiliation{Department of Chemical Engineering, University College London, London WC1E 7JE, United Kingdom}
\email{m.sega@ucl.ac.uk}

\date{\today}

\begin{abstract}

The response of Newtonian liquids to small perturbations is usually considered to be fully described 
by homogeneous transport coefficients like shear and dilatational viscosity. However, the presence of strong density gradients at the liquid/vapor boundary of fluids hints at the possible existence of an  inhomogeneous viscosity. Here, we show that a surface viscosity emerges from the collective dynamics of interfacial layers in molecular simulations of simple liquids. We estimate the surface viscosity to be 8-16 times smaller than that of the bulk fluid at the thermodynamic point considered. This result can have important implications for reactions at liquid surfaces in atmospheric chemistry and catalysis.
\end{abstract}

\maketitle

\section{Introduction}
\begin{figure}[t!]
    \centering
    \includegraphics[width=\columnwidth]{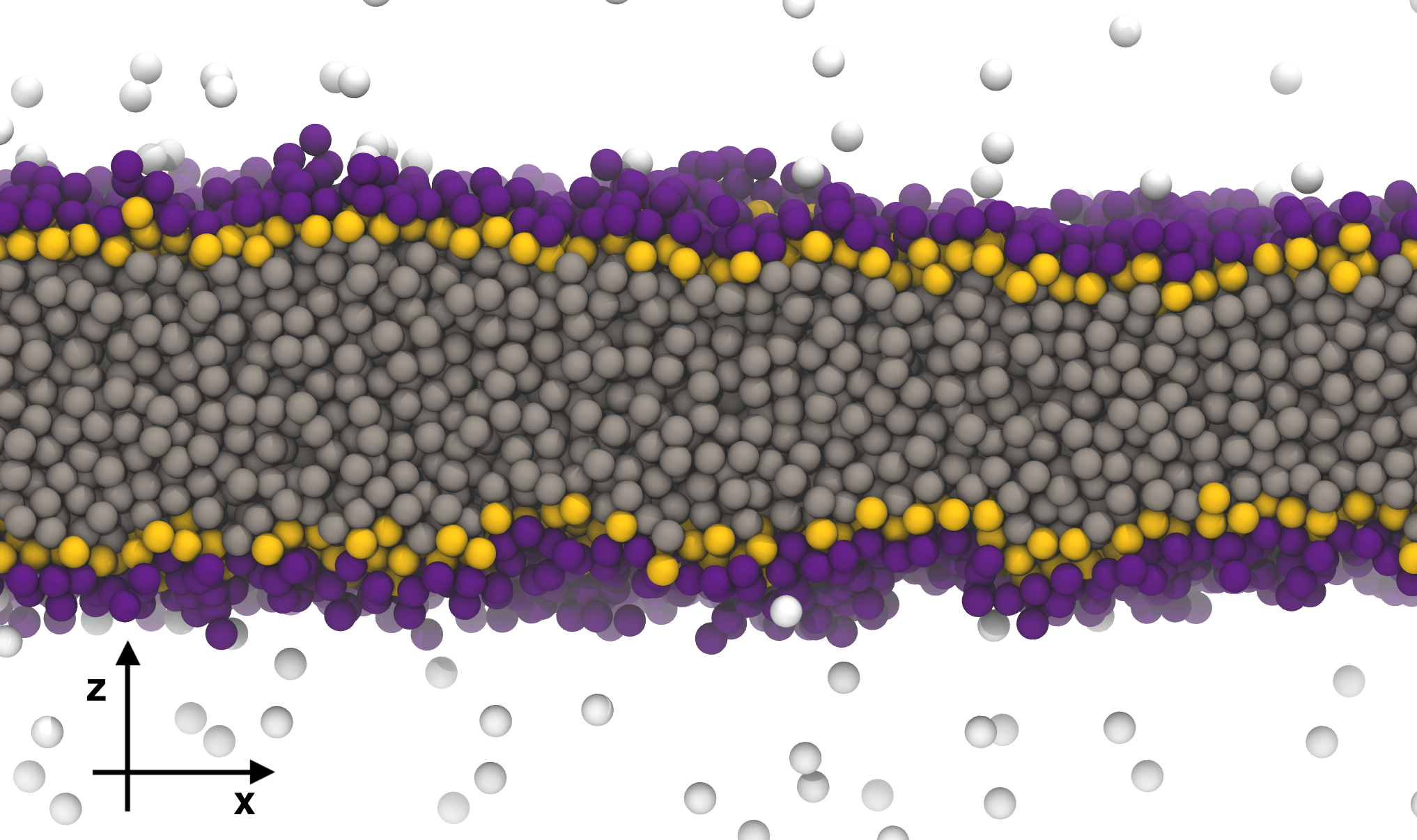}
    \caption{Detail of the liquid/vapor interface in slab configuration (cut through the $x-z$ plane, surface normal along $z$). Light gray: vapor; purple: first layer; yellow: second layer ; dark gray: other liquid phase atoms.}
    \label{fig:snap}
\end{figure}

The concept of surface (shear or dilatational) viscosity has a long history, dating back to Marangoni and Boussinesq\cite{sternling_interfacial_1959,boussinesq_sur_1913,scriven_dynamics_1960} but is usually connected to fluids with complex rheology, where it is a key transport property in processes like liquid film formation\cite{scheid_role_2010}, droplet deformation and breakup\cite{stone_effects_1990}, foam drainage\cite{stone_perspectives_2003,koehler_drainage_2002} cell transport\cite{guglietta_effects_2020}, or electroosmosis\cite{uematsu_power-law_2017}. The possibility of observing a surface viscosity in Newtonian fluids like water was briefly considered in the 1980s\cite{goodrich_theory_1981}. However, initial experimental evidence\cite{earnshaw_surface_1981} was later dismissed\cite{earnshaw_high-frequency_1991}, the rationale behind being that the employed experimental techniques probed micrometer-sized interfacial regions, whereas any possible effect is expected at the molecular scale. Recently, computer simulations gave some indication that this might indeed be the case, as the single particle diffusion coefficient of interfacial molecules can be up to four times larger than in bulk\cite{fabian_single_2017}. The high surface mobility hints at the possible existence of markedly different surface viscosity in simple liquids but calls for a more direct measurement. The presence of a reduced surface viscosity could have profound implications for our understanding of reactions at liquid surfaces\cite{ruiz-lopez_molecular_2020}, in particular for diffusion-limited reactions, as the corresponding acceleration would be unrelated to the presence of hydrogen bonding or dipolar effects\cite{narayan_water_2005,jung_theory_2007,
yan_organic_2016}.
Here, we estimate the viscosity of the surface layers of liquid argon  (see Fig.~\ref{fig:snap}) by comparing the capillary waves  dispersion law obtained from molecular dynamics simulations data with the numerical solutions of the linearized compressible Navier-Stokes equations in presence of a free boundary surface. The results show that the propagation of surface modes is compatible with the presence of modes with a very low surface viscosity, about one tenth of the bulk one, and that bulk-like acoustic modes develop as soon as the second interfacial layer is reached.

\section{Methods}
We model the liquid/vapor interface of argon using the Lennard-Jones potential $U(r) = 4 \epsilon \left[ (\sigma/r)^{12} - (\sigma/r)^6 \right]$ with the parameters reported in Tab.~\ref{table}. The characteristic time scale is given by $\tau = \sigma/\sqrt{\epsilon/m}$. The slab configuration with $N=9807$ atoms is arranged in a $L_x\times{}L_y\times{}L_z$= $18\times{}6\times{}15$ nm$^3$ simulation cell with periodic boundary conditions and interface normal along $z$. The width of the liquid slab is about 4.7 nm, and vapor fills the remaining part of the simulation box.
We simulated the system for 54 ns, storing the configurations to disk for the analysis. Further technical details are reported in the Supplementary Material text, including details on all the algorithms employed\cite{boda_isochoric-_1996,essmann_smooth_1995,nose_molecular_1984,hoover_canonical_1985,michaudagrawal_mdanalysis_2011,sega_phase_2017,rahman_current_1968, boon_molecular_1980,fabian_single_2017,levenberg_method_1944,marquardt_algorithm_1963,falk_capillary_2011,hansen2013theory,savitzky_smoothing_1964}.  The values of the physical quantities appearing in this work (e.g., $\gamma$, $\eta$, $\eta_b$, $c$) are those of liquid argon obtained from the simulation and reported in Tab.~\ref{table}. In the Supplementary Material text, we provide detail on how we calculated these quantities in the slab system, because the procedure is slightly different from the bulk case.

\begin{table}
\centering
\caption{\label{table}Summary of model parameters and physical properties}
    \begin{tabular}{crr}
    \hline
       \textrm{quantity} & \textrm{units} & \multicolumn{1}{c}{\textrm{reduced units}}\\
    \hline
    $\epsilon$ & 0.99641   kJ/mol     & 1 \\
    $\sigma$ & 0.34        nm         & 1 \\
    $m$      & 39.948      amu        & 1 \\
    $\tau$   & 2.1489506   ps         & 1 \\
    $T$      & 93          K          & 0.776 \\
    $p_\textrm{vap}$&1.44(1) bar     &  0.00238(1) \\ 
    $\rho$   & 34.0(1)    mol/L     &  0.805(3) \\
    $\eta$   & 227.0(5)    $\mu$Pa s &  2.497(4) \\ 
    $\eta_b$    & 66.1(3)  $\mu$Pa s &  0.727(4) \\
    $\gamma$ & 13.98(4)    mN/m      &  0.977(6)\\
    $c$      & 790(30)     m/s       &  5.0(2) \\
    \hline
    \end{tabular}
    
\end{table}

\section{Results and Discussion}

To estimate the effective surface viscosity from the molecular dynamics simulations, we compare the molecular dynamics results for the capillary waves dispersion law with the analytical result from continuum hydrodynamics. In the framework of linearized hydrodynamics, the dispersion law for the modes of wavevector  $q=|\mathbf{q}|$ and frequency $\omega$ of a viscous, compressible but thermally nonconducting fluid in presence of a free boundary surface is determined\cite{loudon_theory_1980,harden_hydrodynamic_1991,mora_height_2002,falk_capillary_2011} by the zeros of 
\begin{equation}
  D(q,\omega) = -\frac{\gamma q^2}{\rho_m} q_\rho + \omega^2+ \frac{4i\omega}{\rho_m} \eta q^2 - 4 \frac{\eta^2}{\rho^2_m} q^2(q^2-q_vq_\rho).\label{eq:dispersion}
\end{equation}
Here, $\rho_m = m\rho$, with $\rho$ the number density and $m$ the molecular mass, $\gamma$ is the surface tension, $\eta$ the dynamic shear viscosity, 
\begin{equation} q_v^2 = q^2 - i \omega \rho_m /\eta,\end{equation} and  
\begin{equation}(q^2 - q_\rho^2)[c^2/g-  i\omega(\eta_b+4\eta/3)/\rho_m] = \omega^2.\end{equation} 
The heat capacity ratio is $g=c_p/c_v$ is and $\eta_b$ is the bulk or dilatational viscosity. Note that in literature the completely equivalent form that uses a different definition of the Fourier transform is often used, and can be recovered by changing the sign of $\omega$ in Eq.~\ref{eq:dispersion}. 
For simple liquids far from the critical point, the deviations from the conducting case are minimal\cite{loudon_theory_1980}. Therefore, one can safely disregard the presence of thermal conductivity.

Away from the hydrodynamic limit of vanishing $\omega$ and $q$, the perturbations sooner or later reach the characteristic length and time scales of molecular processes. If these changes are smooth enough, it is possible to extend the description to include molecular correlations while retaining the general structure of hydrodynamics\cite{boon_molecular_1980} but introducing, for example, wavevector-dependent transport properties. This is the case for the wavevector-dependent shear viscosity $\eta(q)$ in bulk liquids\cite{rahman_current_1968,chung_generalized_1969,gaskell_wavevector-dependent_1987} or surface tension $\gamma(q)$ in interfacial systems\cite{mecke_effective_1999,hofling_enhanced_2015,delgado-buscalioni_hydrodynamics_2008,hernandez-munoz_layering_2022}.

Here, we do not concern ourselves with the region of large wavevectors, but are interested in the macroscopic, hydrodynamic limit of $q\to 0$. In bulk simple liquids, the effects of correlations do not show up in the dispersion before $q\sigma \simeq 2$\cite{rahman_current_1968,boon_molecular_1980}, where $\sigma$ identifies the molecular diameter. At lower values of $q$, it is usually safe to interpret the results of molecular simulations in terms of continuum hydrodynamics, and our analysis is restricted to this regime, even though, for completeness, we show also the behavior at larger wavevectors.

Further, it should be noted that at small scales nonlocal effects can arise\cite{hansen2007parameterization,duque2020non}. Even though we are considering the $q\to 0$ limit along the surface plane, the strong anisotropy along $z$ implies that, in principle, close to the surface one cannot neglect nonlocality. In fact, it should be kept in mind that by using a continuum model of hydrodynamics with constant (and local) transport coefficients to fit the simulation results, we are indeed calculating an effective viscosity.

\begin{figure}[t!]
    \centering
    \includegraphics[width=\columnwidth,clip,trim=5 5 30 30]{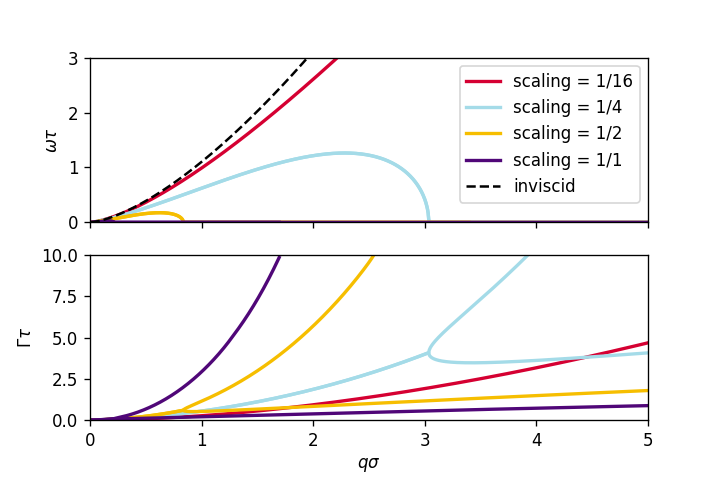}    
    \includegraphics[width=\columnwidth,clip,trim=5 5 30 30]{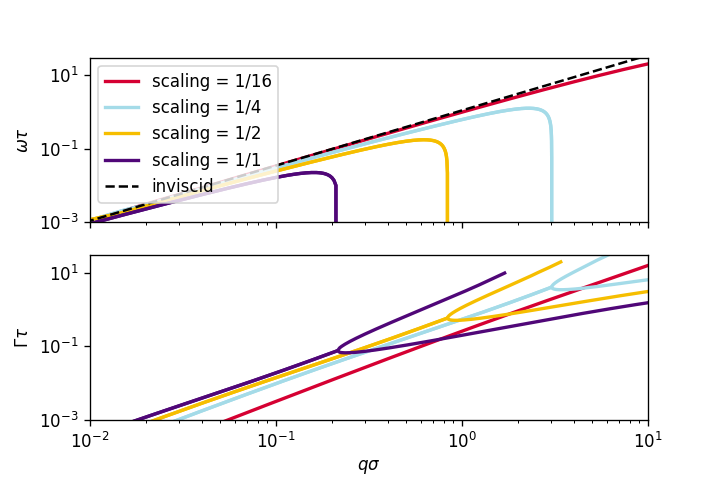}    
    \caption{Real ($\omega\tau$) and imaginary ($\Gamma \tau$) part of the dispersion law $\omega(q)$ obtained from the numerical solutios of the zeros of Eq.~\ref{eq:dispersion}. The physical parameters correspond to those reported in Tab.~\ref{table}, with a scaling factor (1,1/2, 1/4, 1/16) applied to the shear and dilatational viscosities. Top two panels: linear scale; bottom two panels: logarithmic scale. The code for the calculation of the dispersion curves is available on Zenodo\cite{zenodo7416368}.
    \label{fig:dispersion}}
\end{figure}

The first step that is needed to interpret any results of molecular simulations in the low-$q$ region is to obtain the solution of the linearized hydrodynamic equations, which allows to compare the particle-based simulation results with the continuum theory. We solved Eq.~\ref{eq:dispersion} numerically and obtained a family of dispersion curves by rescaling homogeneously the shear and dilatational viscosity from the value of the atomistic model of liquid argon towards zero. All other physical parameters were kept fixed at the corresponding values of the atomistic model, reported in Tab.~\ref{table}. The dispersion curves are shown in Fig.~\ref{fig:dispersion} in linear and double-logarithmic scale. All curves' real parts follow  an initial growth that is similar to the inviscid case, $\omega(q)= q^{3/2} \sqrt{\gamma/\rho_m}$, and then drop abruptly to zero at a critical value $q_c(\eta)$. The vanishing of the real part of the dispersion law means that for $q>q_c$, signals do not propagate anymore. In correspondence of $q_c$, the imaginary part of the dispersion curve has a branching point\cite{falk_capillary_2011}, after which it splits into two. Note that the ideal solution of the inviscid fluid has no absorption. 
In addition to these solutions, non-hydrodynamic branches can develop at higher values of $q$. Here, we disregard these solutions as we are interested only in those that reach the macroscopic, hydrodynamic limit $q=0$ and in all figures  we report only the branches stemming from the origin.

Obtaining the numerical solutions of the dispersion law as presented in  Fig.~\ref{fig:dispersion} represents the first step for the interpretation of the microscopic model obtained by molecular dynamics simulations. To access the dispersion law $\omega(q)$ of the surface layer from the molecular dynamic simulations, we looked at the spectra of the hydrodynamic modes fluctuations\cite{boon_molecular_1980}, restricting their calculation to  the set of atoms belonging to the surface layer.  To select a mode, we choose the associated wavevector $\mathbf{q}$ as pointing along $x$ and compute the time series of the collective currents 
\begin{equation} J_k^\alpha(t) = \sum_i v^\alpha_i \exp(iq_kx_i)\label{eq:j}\end{equation}  and height modes \begin{equation}h_k(t) =\sum_i z_i \exp(iq_kx_i),\label{eq:h}\end{equation} where $\alpha=x,y $ or $z$. 
The index $i$ labels the atoms belonging to either the surface or the second layer. The wavevectors compatible with the periodic boundary conditions are in  the form $q_k = 2\pi k / L_x$, with $k\in\{1,2,\ldots\}$.
The molecular layers are determined using the Identification of Truly Interfacial Molecules (ITIM) analysis\cite{partay_new_2008,sega_pytim_2018}. This technique has already proven helpful to bridge between the molecular and continuum description of free surface hydrodynamics\cite{giri_resolving_2022}. In a nutshell, the method identifies surface atoms as those which are exposed to the vapor phase, thereby taking into account the fluctuations induced by thermal capillary waves. Once all the atoms in the first layer are identified, the same procedure can be applied to the successive molecular layer. More detail is available in the Supplementary Information (Supplementary Material) text.

Unlike single particle dynamical properties such as the diffusion coefficient, collective ones like the currents, Eqs.~\ref{eq:j} and \ref{eq:h} can be defined without problem on a set of particles that changes with time, so  even if particles are leaving and joining the layer. This is the core idea behind the approach that we propose here and that allows to compute collective properties for a subset of the atoms in the simulation box, thereby giving access, in the present case, to the position-dependent equivalent of the shear viscosity, which we can now calculate on a layer-by-layer basis.

 From the time series $J_k^\alpha(t) $ and $h_k(t)$ we compute the autocorrelation functions \begin{equation}C_J^\alpha(q_k,t)  =\left\langle J_k^\alpha(0) J_k^\alpha(t)\right\rangle,\end{equation}  \begin{equation} C_h(q_k,t)  = \langle h_k(0)h_k(t)\rangle,\end{equation}  and the respective spectra 
\begin{equation}
    \tilde{C}_J^\alpha(q_k,\omega)  = \int_{-\infty}^\infty dt \exp(i\omega t) C_J^\alpha(q_k,t),
\end{equation}
\begin{equation}
    \tilde{C}_h(q_k,\omega)  = \int_{-\infty}^\infty dt \exp(i\omega t) C_h(q_k,t).
\end{equation}
Note that in bulk hydrodynamics, the symmetry of the problem reduces the number of currents and the respective modes to two: a longitudinal one ($\mathbf{q}\cdot\mathbf{v}=0$) and two degenerate transverse modes ($\mathbf{q}\times\mathbf{v}=0$). Bulk longitudinal and transverse modes decouple, and the latter, in continuum hydrodynamics, cannot propagate. The presence of the boundary surface (normal along $z$) removes this degeneracy, and three types of collective current modes emerge. Two, $C_J^x$ and $C_J^z$, are coupled. The third one, $C_J^y$, is decoupled from the previous two and, in continuum hydrodynamics, has a purely imaginary dispersion relation, i.e., no surface shear modes can exist\cite{loudon_theory_1980}.
Sample spectra are reported in the Supplementary Material figures S1-S5. Here, to characterize the real part of the dispersion relation $\omega(q)$ we computed the frequency of the maximum of the collective currents spectra, $\omega_\mathrm{max}(q)$, a typical proxy for this quantity\cite{boon_molecular_1980}. 

\begin{figure}[t!]
    \centering
    \includegraphics[width=\columnwidth,clip,trim=20 5 40 30]{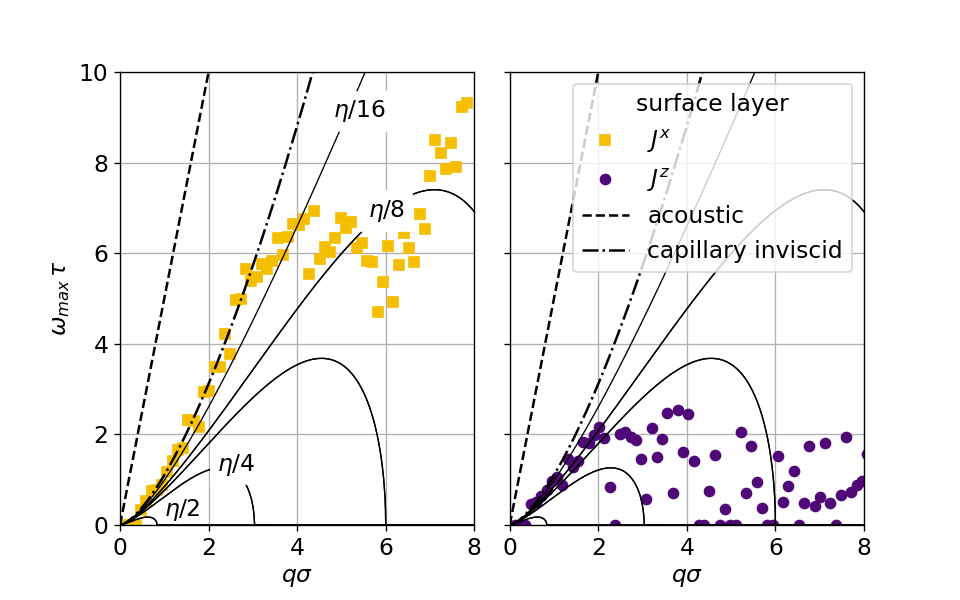}    
    \includegraphics[width=\columnwidth,clip,trim=20 5 40 30]{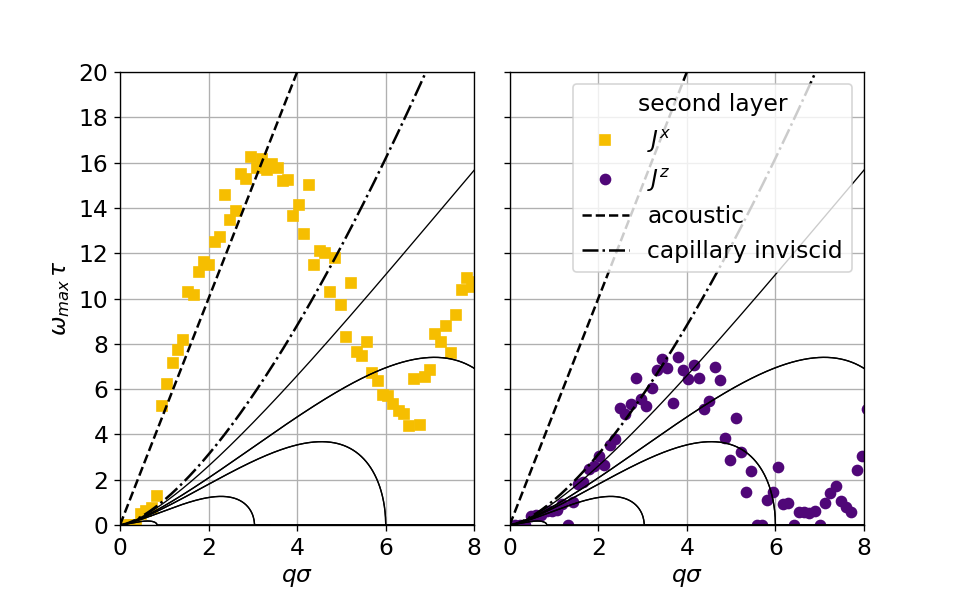}    
    \caption{$\omega_\mathrm{max}$ for the spectra of $C_J^x(\omega,q)$ (left column) and $C_J^z(\omega,q)$ (right column) for the surface (top row) and second  molecular layer (bottom row). The dashed and dot-dashed curves are the acoustic ($\omega = c q)$ and capillary ($\omega = q^{3/2}\sqrt{\gamma/\rho_m}$) modes. The solid lines are the real part of the dispersion law  computed numerically for different viscosity values.}
    \label{fig:omega}
\end{figure}

In Fig.~\ref{fig:omega}, we report $\omega_\mathrm{max}(q)$ of the first two molecular layers for components $x$ and $z$.  To interpret the propagating modes, we compare the molecular dynamics results to the analytical solutions 
of $D(q,\omega)=0$ for different choices of the viscosity parameter expressed as fractions of $\eta$ ($\eta_b$ being rescaled accordingly).  These curves are reported in Fig.~\ref{fig:omega} (thin solid lines), along with the sound dispersion in bulk $\omega = c q$  (dashed line) and the the capillary dispersion of an ideal, inviscid fluid, $\omega = q^{3/2} \sqrt{\gamma/\rho_m}$ (dot-dashed line).
Before discussing the results on the propagating modes $C_J^x$ and $C_J^z$, we note that the remaining two, $C_J^y$ and $C_h$ are non-propagating, that is,  for them $\omega_\mathrm{max}(q)=0$ , as shown  in Fig.~3 of the Supplementary Material. The absence of propagation for $C_J^y$ in the hydrodynamic regime is expected, as it is a transverse mode\cite{boon_molecular_1980}. Interestingly, we observe no propagation also at high values of $q$, differently from the bulk case\cite{rahman_current_1968}. The absence of propagation for $C_h$, on the other hand, confirms the results of Ref.~\cite{delgado-buscalioni_hydrodynamics_2008}.

The emerging picture from the $C_J^x$ and $C_J^z$ spectra of the first layer reported in Fig.\ref{fig:omega} is that of a dispersion law compatible with a very low viscosity, in the range from $\eta/16$ to $\eta/8$, at least up to $q\sigma \simeq 2$, as it is clearly recognizable in the top row of Fig.~\ref{fig:omega} when comparing the molecular dynamics results with the numerical solution of the hydrodynamic theory. At values of $q\sigma$ larger than 2, continuum hydrodynamics starts breaking down and it would be pointless to compare the solutions with the molecular dynamics results. In the second layer (bottom row of Fig.~\ref{fig:omega}) there are still traces of a capillary mode along $x$ (left panel), but the main peak quickly shifts along the acoustic branch at $q\sigma\simeq{}1$. This result suggests that sound waves can propagate (parallel to the surface) as soon as molecules are just below the surface layer. This is a remarkable demonstration of the sharpness of the interface, showing that bulk-like characteristics appear already in the second molecular layer not only for structural properties\cite{chacon_intrinsic_2003,sega_layer-by-layer_2015}, but also for collective transport ones.

Overall, the collective dynamics appears to be enhanced by a factor not too dissimilar from that observed for single particle diffusion within the surface layer of carbon tetrachloride\cite{fabian_single_2017} which, like argon, is a good model of a simple liquid. 
Still, the results reported in Fig.~\ref{fig:omega} do not rule out the presence of a capillary mode with a viscosity equal to that of the bulk fluid, $\eta$. The value of $q_c$ decreases with increasing viscosity, and the numerical solution of Eq.~\ref{eq:dispersion} using  $\eta$ as viscosity yields $q_c\sigma \simeq0.22$, a value so small that is practically impossible to resolve accurately with the current simulation, despite the long sampling. In this case, one would not be able to observe any propagating mode from the simulation data. This is exactly what happens for the modes of $h$. The real part of the dispersion curve for $C_h$ is, in this sense, compatible with a viscosity equal to that of the bulk fluid.

This is only an apparent conundrum. It is not surprising that different correlation functions underline the presence of some modes and hide other, even though they share the same dispersion law. This happens in the bulk fluid, where the Rayleigh peak associated to the thermal diffusion appears clearly in the dynamic structure factor, but it is suppressed in the longitudinal current\cite{boon_molecular_1980} due to the presence of a $\omega^2$ factor linking the two. 
A similar picture seems to emerge here, with modes that are compatible with a small viscosity being clearly visible in  the spectrum of the longitudinal current, and other with a larger viscosity being highlighted in the spectrum of the surface height. After all, the viscosity of the bulk fluid has to show its influence on the capillary waves, because the dispersion law of the surface has to transform into that of the bulk when penetrating into the liquid (we see the appearance of the sound dispersion law in the second layer already), and the dynamics of subsequent layers are coupled through the interatomic interactions. In this sense, the system behaves as two coupled fluids (a thin interfacial layer and the underlying bulk) with different viscosities.

\begin{figure}[t!]
    \centering
    \includegraphics[width=\columnwidth,clip,trim=10 5 40 30]{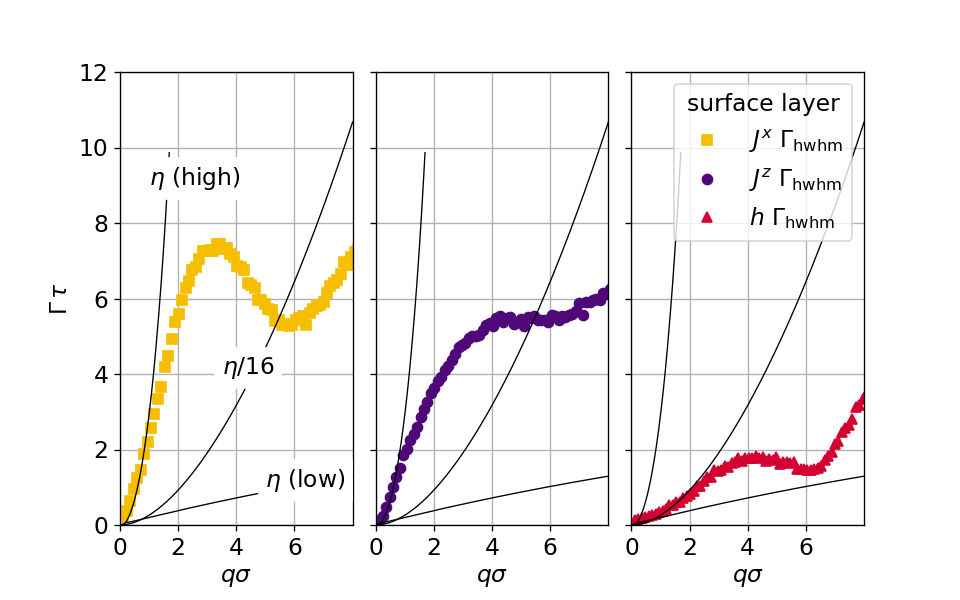}    
    \includegraphics[width=\columnwidth,clip,trim=10 5 40 30]{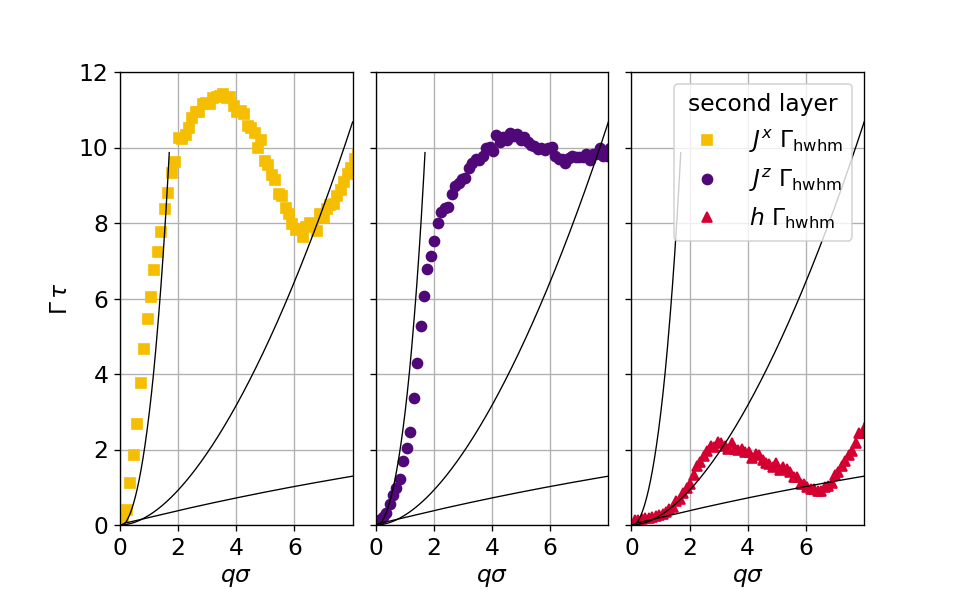}    
    \caption{Imaginary part of the dispersion curve estimated from $C_J^x$ (left column),  $C_J^z$ (central column)  and $C_h$ (right column), for the surface  (top row) and second molecular layer (bottom row).  Filled symbols: $\Gamma_\mathrm{min}$; open symbols: $\Gamma_\mathrm{hwhm}$.  The solid lines are the imaginary part of the numerical solution of $D(q,\omega)=0$. }
    \label{fig:gamma}
\end{figure}

Looking at the imaginary part of the dispersion law (the absorption) provides further insight.   For a single Lorentzian spectrum $L(\omega) \propto 1/[ (\omega - \omega_\mathrm{max})^2 + \Gamma^2] $, the line broadening $\Gamma$ is the curve's half width at half maximum (hwhm), and determines the inverse time of a perturbation's exponential decay. However, the present spectra can be described reasonably well (see Figs.~S1-S5 of the Supplementary Material) only by a superposition of three Lorentzian functions. To avoid any bias induced by the fitting procedure, we decided to use the hwhm of the whole spectrum, $\Gamma_\mathrm{hwhm}$, as a proxy for the absorption. The meaning of this quantity is not as direct as that of a single Lorentzian spectral line. However, $\Gamma_\mathrm{hwhm}$ is a quite stable quantity, like $\omega_\mathrm{max}$, against variation in the analysis protocol (e.g., the length of the correlation function used to compute the spectra, or the filtering window size discussed in the Supplementary Material text) and does not depend on a fitting procedure. 

In Fig.~\ref{fig:gamma}, we report the values of $\Gamma_\mathrm{hwhm}$,  computed for the first two molecular layers, as well as the imaginary part of the zeros of $D(q,\omega)$ calculated numerically, for viscosity values $\eta$ and $\eta/16$ (same scaling factor applied to the bulk viscosity). Notice that for a viscosity value $\eta$ the imaginary part branches off\cite{falk_capillary_2011}  in correspondence to the critical wavevector $q_c\sigma\simeq0.22$ into two solutions. One solution (low) shows a slowly increasing, roughly linear function of $q$, while the other (high) is a rapidly growing function of $q$. For the lower viscosity case, $\eta/16$, $q_c$ lies outside the visible range of $q$, and only one curve is seen in the plots.  In both the spectra of $C_J^x$ and $C_J^z$, $\Gamma_\mathrm{hwhm}$ shows a steep rise along the high branch for viscosity $\eta$.  The spectrum of $C_h$, which is always peaked around zero, shows instead a width $\Gamma_\mathrm{hwhm}$ that seems to follow the $\eta/16$ dispersion curve relatively well. The absorption data support the same picture that emerges from the analysis of $\omega_\mathrm{max}$, showing again that two families of modes appear in the surface layers, one of which can be described by a viscosity that is markedly lower than that of the bulk fluid. 

Here, a further consideration is in order regarding the non-evaporation boundary condition, $J_z = \partial h / \partial t$, typically used in the hydrodynamic theory (e.g., in the derivation of Eq.~\ref{eq:dispersion}). This condition does not, strictly speaking,  apply to the present case, not because of evaporating molecules (present, but relatively rare) but because of the way the surface layer is defined. In the present formulation, atoms can join/leave the surface layer from/to the layer underneath, thereby changing both $J_z$ and $h$ in an impulsive way. This makes, by the way, the derivative of the correlation functions different from zero at $t=0$, a condition otherwise imposed by time reversal\cite{hansen2013theory}. As a consequence, one cannot expect the respective spectra to be related by $C_J^z(q, \omega) = -i\omega C_h(q,\omega)$.

\section*{Conclusions}
The early intuition of Goodrich\cite{goodrich_theory_1981} about the existence of surface excess viscosity in simple liquids, and the assumption of Earnshaw\cite{earnshaw_high-frequency_1991} about their hypothetical confinement in a thin molecular region turned out to be, in essence, correct,  even though there is no need to invoke the presence of a diffuse interface. In fact, the strong anisotropy and inhomogeneity at the sharp liquid/vapor interface promotes a faster collective dynamics of the surface layer with respect to the bulk and, in turn, a much lower viscosity associated to some of its collective modes.

Continuum hydrodynamics is known to break down in simple liquids at scales smaller than about three molecular diameters, where correlations and nonlocal effects start playing an important role. Despite the strong anisotropy of the liquid/vapor interface, effective transport properties can still be computed in the long wavelength limit.
The collective currents in the first layer show the unmistakable signature of capillary modes, while acoustic modes can start propagating along the surface already in the second molecular layer, showing that bulk-like properties start appearing as soon as molecules enter the second surface layer.
We interpreted the spectra of the collective currents in terms of two families of modes, the first associated with the viscosity of the bulk fluid and the second with a much smaller surface viscosity. We relate the coexistence of these two families of modes to the presence of a more mobile set of surface atoms, a finding backed by results on the diffusion coefficient of surface molecules. To which extent the low viscosity modes extend deep in the $q\to 0$ limit is an open question.  

The presence of a reduced effective viscosity on the surface can be ascribed to the increased mobility of interfacial atoms, thanks to the asymmetric environment. One might wonder what can happen at liquid/solid or liquid/liquid interfaces. A strong interaction with a rigid substrate is likely to reduce the mobility\cite{sendner_interfacial_2009} and induce an increase in the surface viscosity. One the other hand, the presence of slip at solid surfaces\cite{thompson1990shear,hansen2011prediction} (albeit very smooth ones\cite{ho2011liquid,sega2013regularization,sega2015importance}) and liquid/liquid interfaces\cite{hilaire2023liquid} signals a weaker interaction with the opposite phase and could be compatible with a reduction in effective surface viscosity, since the slip length is essentially a measure of the ratio between viscosity and frictional forces.

For a free liquid surface, the measured dispersion relations support the existence of a surface effective viscosity in simple liquids that is about one order of magnitude smaller than in the bulk. This finding could have important implications for the kinetics of diffusion-limited reactions at liquid interfaces. 
Given the role of viscosity in reaction kinetics\cite{eyring_viscosity_1936} and the importance of reactions at liquid surfaces\cite{benjamin_chemical_1996} in atmospheric chemistry\cite{finlayson-pitts_reactions_2009} or catalysis\cite{narayan_water_2005,jung_theory_2007,taccardi_gallium-rich_2017}, it is hard to underestimate the impact of such a  small viscosity at the surface of simple liquids, even if limited to the range of wavevectors explored in this work. This would imply an acceleration of diffusion-limited reactions by a factor inversely proportional to the viscosity ratio, therefore, in the range 8-16, a trend opposite to that observed at liquid/solid interfaces\cite{sendner_interfacial_2009}.

\section*{Supplementary material}
See the supplementary material for additional information on the simulations; details on the Green-Kubo calculation of transport coefficients in inhomogeneous systems; details on the algorithm for the identification of surface molecules; and plots of selected autocorrelation functions and relative spectra.

\bibliography{references}

\begin{thebibliography}{58}%
\makeatletter
\providecommand \@ifxundefined [1]{%
 \@ifx{#1\undefined}
}%
\providecommand \@ifnum [1]{%
 \ifnum #1\expandafter \@firstoftwo
 \else \expandafter \@secondoftwo
 \fi
}%
\providecommand \@ifx [1]{%
 \ifx #1\expandafter \@firstoftwo
 \else \expandafter \@secondoftwo
 \fi
}%
\providecommand \natexlab [1]{#1}%
\providecommand \enquote  [1]{``#1''}%
\providecommand \bibnamefont  [1]{#1}%
\providecommand \bibfnamefont [1]{#1}%
\providecommand \citenamefont [1]{#1}%
\providecommand \href@noop [0]{\@secondoftwo}%
\providecommand \href [0]{\begingroup \@sanitize@url \@href}%
\providecommand \@href[1]{\@@startlink{#1}\@@href}%
\providecommand \@@href[1]{\endgroup#1\@@endlink}%
\providecommand \@sanitize@url [0]{\catcode `\\12\catcode `\$12\catcode
  `\&12\catcode `\#12\catcode `\^12\catcode `\_12\catcode `\%12\relax}%
\providecommand \@@startlink[1]{}%
\providecommand \@@endlink[0]{}%
\providecommand \url  [0]{\begingroup\@sanitize@url \@url }%
\providecommand \@url [1]{\endgroup\@href {#1}{\urlprefix }}%
\providecommand \urlprefix  [0]{URL }%
\providecommand \Eprint [0]{\href }%
\providecommand \doibase [0]{http://dx.doi.org/}%
\providecommand \selectlanguage [0]{\@gobble}%
\providecommand \bibinfo  [0]{\@secondoftwo}%
\providecommand \bibfield  [0]{\@secondoftwo}%
\providecommand \translation [1]{[#1]}%
\providecommand \BibitemOpen [0]{}%
\providecommand \bibitemStop [0]{}%
\providecommand \bibitemNoStop [0]{.\EOS\space}%
\providecommand \EOS [0]{\spacefactor3000\relax}%
\providecommand \BibitemShut  [1]{\csname bibitem#1\endcsname}%
\let\auto@bib@innerbib\@empty
\bibitem [{\citenamefont {Sternling}\ and\ \citenamefont
  {Scriven}(1959)}]{sternling_interfacial_1959}%
  \BibitemOpen
  \bibfield  {author} {\bibinfo {author} {\bibfnamefont {C.~V.}\ \bibnamefont
  {Sternling}}\ and\ \bibinfo {author} {\bibfnamefont {L.~E.}\ \bibnamefont
  {Scriven}},\ }\bibfield  {title} {\enquote {\bibinfo {title} {Interfacial
  turbulence: Hydrodynamic instability and the {{Marangoni}} effect},}\ }\href
  {\doibase 10.1002/aic.690050421} {\bibfield  {journal} {\bibinfo  {journal}
  {AIChE Journal}\ }\textbf {\bibinfo {volume} {5}},\ \bibinfo {pages}
  {514--523} (\bibinfo {year} {1959})}\BibitemShut {NoStop}%
\bibitem [{\citenamefont {Boussinesq}(1913)}]{boussinesq_sur_1913}%
  \BibitemOpen
  \bibfield  {author} {\bibinfo {author} {\bibfnamefont {M.~J.}\ \bibnamefont
  {Boussinesq}},\ }\bibfield  {title} {\enquote {\bibinfo {title} {Sur
  l'existence d'une viscosit\'e superficielle, dans la mince couche de
  transition s\'eparant un liquide d'un autre fluide contigu},}\ }\href@noop {}
  {\bibfield  {journal} {\bibinfo  {journal} {Ann. Chim. Phys}\ }\textbf
  {\bibinfo {volume} {29}},\ \bibinfo {pages} {349--357} (\bibinfo {year}
  {1913})}\BibitemShut {NoStop}%
\bibitem [{\citenamefont {Scriven}(1960)}]{scriven_dynamics_1960}%
  \BibitemOpen
  \bibfield  {author} {\bibinfo {author} {\bibfnamefont {L.~E.}\ \bibnamefont
  {Scriven}},\ }\bibfield  {title} {\enquote {\bibinfo {title} {Dynamics of a
  fluid interface equation of motion for {{Newtonian}} surface fluids},}\
  }\href@noop {} {\bibfield  {journal} {\bibinfo  {journal} {Chemical
  Engineering Science}\ }\textbf {\bibinfo {volume} {12}},\ \bibinfo {pages}
  {98--108} (\bibinfo {year} {1960})}\BibitemShut {NoStop}%
\bibitem [{\citenamefont {Scheid}\ \emph {et~al.}(2010)\citenamefont {Scheid},
  \citenamefont {Delacotte}, \citenamefont {Dollet}, \citenamefont {Rio},
  \citenamefont {Restagno}, \citenamefont {{van Nierop}}, \citenamefont
  {Cantat}, \citenamefont {Langevin},\ and\ \citenamefont
  {Stone}}]{scheid_role_2010}%
  \BibitemOpen
  \bibfield  {author} {\bibinfo {author} {\bibfnamefont {B.}~\bibnamefont
  {Scheid}}, \bibinfo {author} {\bibfnamefont {J.}~\bibnamefont {Delacotte}},
  \bibinfo {author} {\bibfnamefont {B.}~\bibnamefont {Dollet}}, \bibinfo
  {author} {\bibfnamefont {E.}~\bibnamefont {Rio}}, \bibinfo {author}
  {\bibfnamefont {F.}~\bibnamefont {Restagno}}, \bibinfo {author}
  {\bibfnamefont {E.~A.}\ \bibnamefont {{van Nierop}}}, \bibinfo {author}
  {\bibfnamefont {I.}~\bibnamefont {Cantat}}, \bibinfo {author} {\bibfnamefont
  {D.}~\bibnamefont {Langevin}}, \ and\ \bibinfo {author} {\bibfnamefont
  {H.~A.}\ \bibnamefont {Stone}},\ }\bibfield  {title} {\enquote {\bibinfo
  {title} {The role of surface rheology in liquid film formation},}\ }\href
  {\doibase 10.1209/0295-5075/90/24002} {\bibfield  {journal} {\bibinfo
  {journal} {EPL}\ }\textbf {\bibinfo {volume} {90}},\ \bibinfo {pages} {24002}
  (\bibinfo {year} {2010})}\BibitemShut {NoStop}%
\bibitem [{\citenamefont {Stone}\ and\ \citenamefont
  {Leal}(1990)}]{stone_effects_1990}%
  \BibitemOpen
  \bibfield  {author} {\bibinfo {author} {\bibfnamefont {H.~A.}\ \bibnamefont
  {Stone}}\ and\ \bibinfo {author} {\bibfnamefont {L.~G.}\ \bibnamefont
  {Leal}},\ }\bibfield  {title} {\enquote {\bibinfo {title} {The effects of
  surfactants on drop deformation and breakup},}\ }\href@noop {} {\bibfield
  {journal} {\bibinfo  {journal} {Journal of Fluid Mechanics}\ }\textbf
  {\bibinfo {volume} {220}},\ \bibinfo {pages} {161--186} (\bibinfo {year}
  {1990})}\BibitemShut {NoStop}%
\bibitem [{\citenamefont {Stone}\ \emph {et~al.}(2003)\citenamefont {Stone},
  \citenamefont {Koehler}, \citenamefont {Hilgenfeldt},\ and\ \citenamefont
  {Durand}}]{stone_perspectives_2003}%
  \BibitemOpen
  \bibfield  {author} {\bibinfo {author} {\bibfnamefont {H.~A.}\ \bibnamefont
  {Stone}}, \bibinfo {author} {\bibfnamefont {S.~A.}\ \bibnamefont {Koehler}},
  \bibinfo {author} {\bibfnamefont {S.}~\bibnamefont {Hilgenfeldt}}, \ and\
  \bibinfo {author} {\bibfnamefont {M.}~\bibnamefont {Durand}},\ }\bibfield
  {title} {\enquote {\bibinfo {title} {Perspectives on foam drainage and the
  influence of interfacial rheology},}\ }\href {\doibase
  10.1088/0953-8984/15/1/338} {\bibfield  {journal} {\bibinfo  {journal} {J.
  Phys.: Condens. Matter}\ }\textbf {\bibinfo {volume} {15}},\ \bibinfo {pages}
  {S283--S290} (\bibinfo {year} {2003})}\BibitemShut {NoStop}%
\bibitem [{\citenamefont {Koehler}\ \emph {et~al.}(2002)\citenamefont
  {Koehler}, \citenamefont {Hilgenfeldt}, \citenamefont {Weeks},\ and\
  \citenamefont {Stone}}]{koehler_drainage_2002}%
  \BibitemOpen
  \bibfield  {author} {\bibinfo {author} {\bibfnamefont {S.~A.}\ \bibnamefont
  {Koehler}}, \bibinfo {author} {\bibfnamefont {S.}~\bibnamefont
  {Hilgenfeldt}}, \bibinfo {author} {\bibfnamefont {E.~R.}\ \bibnamefont
  {Weeks}}, \ and\ \bibinfo {author} {\bibfnamefont {H.~A.}\ \bibnamefont
  {Stone}},\ }\bibfield  {title} {\enquote {\bibinfo {title} {Drainage of
  single {{Plateau}} borders: {{Direct}} observation of rigid and mobile
  interfaces},}\ }\href {\doibase 10.1103/PhysRevE.66.040601} {\bibfield
  {journal} {\bibinfo  {journal} {Phys. Rev. E}\ }\textbf {\bibinfo {volume}
  {66}},\ \bibinfo {pages} {040601} (\bibinfo {year} {2002})}\BibitemShut
  {NoStop}%
\bibitem [{\citenamefont {Guglietta}\ \emph {et~al.}(2020)\citenamefont
  {Guglietta}, \citenamefont {Behr}, \citenamefont {Biferale}, \citenamefont
  {Falcucci},\ and\ \citenamefont {Sbragaglia}}]{guglietta_effects_2020}%
  \BibitemOpen
  \bibfield  {author} {\bibinfo {author} {\bibfnamefont {F.}~\bibnamefont
  {Guglietta}}, \bibinfo {author} {\bibfnamefont {M.}~\bibnamefont {Behr}},
  \bibinfo {author} {\bibfnamefont {L.}~\bibnamefont {Biferale}}, \bibinfo
  {author} {\bibfnamefont {G.}~\bibnamefont {Falcucci}}, \ and\ \bibinfo
  {author} {\bibfnamefont {M.}~\bibnamefont {Sbragaglia}},\ }\bibfield  {title}
  {\enquote {\bibinfo {title} {On the effects of membrane viscosity on
  transient red blood cell dynamics},}\ }\href@noop {} {\bibfield  {journal}
  {\bibinfo  {journal} {Soft Matter}\ }\textbf {\bibinfo {volume} {16}},\
  \bibinfo {pages} {6191--6205} (\bibinfo {year} {2020})}\BibitemShut {NoStop}%
\bibitem [{\citenamefont {Uematsu}, \citenamefont {Netz},\ and\ \citenamefont
  {Bonthuis}(2017)}]{uematsu_power-law_2017}%
  \BibitemOpen
  \bibfield  {author} {\bibinfo {author} {\bibfnamefont {Y.}~\bibnamefont
  {Uematsu}}, \bibinfo {author} {\bibfnamefont {R.~R.}\ \bibnamefont {Netz}}, \
  and\ \bibinfo {author} {\bibfnamefont {D.~J.}\ \bibnamefont {Bonthuis}},\
  }\bibfield  {title} {\enquote {\bibinfo {title} {Power-law electrokinetic
  behavior as a direct probe of effective surface viscosity},}\ }\href
  {\doibase 10.1016/j.cplett.2016.12.056} {\bibfield  {journal} {\bibinfo
  {journal} {Chemical Physics Letters}\ }\textbf {\bibinfo {volume} {670}},\
  \bibinfo {pages} {11--15} (\bibinfo {year} {2017})}\BibitemShut {NoStop}%
\bibitem [{\citenamefont {Goodrich}(1981)}]{goodrich_theory_1981}%
  \BibitemOpen
  \bibfield  {author} {\bibinfo {author} {\bibfnamefont {F.~C.}\ \bibnamefont
  {Goodrich}},\ }\bibfield  {title} {\enquote {\bibinfo {title} {The theory of
  capillary excess viscosities},}\ }\href@noop {} {\bibfield  {journal}
  {\bibinfo  {journal} {Proceedings of the Royal Society of London. A.
  Mathematical and Physical Sciences}\ }\textbf {\bibinfo {volume} {374}},\
  \bibinfo {pages} {341--370} (\bibinfo {year} {1981})}\BibitemShut {NoStop}%
\bibitem [{\citenamefont {Earnshaw}(1981)}]{earnshaw_surface_1981}%
  \BibitemOpen
  \bibfield  {author} {\bibinfo {author} {\bibfnamefont {J.~C.}\ \bibnamefont
  {Earnshaw}},\ }\bibfield  {title} {\enquote {\bibinfo {title} {Surface
  viscosity of water},}\ }\href {\doibase 10.1038/292138a0} {\bibfield
  {journal} {\bibinfo  {journal} {Nature}\ }\textbf {\bibinfo {volume} {292}},\
  \bibinfo {pages} {138--139} (\bibinfo {year} {1981})}\BibitemShut {NoStop}%
\bibitem [{\citenamefont {Earnshaw}\ and\ \citenamefont
  {Hughes}(1991)}]{earnshaw_high-frequency_1991}%
  \BibitemOpen
  \bibfield  {author} {\bibinfo {author} {\bibfnamefont {J.~C.}\ \bibnamefont
  {Earnshaw}}\ and\ \bibinfo {author} {\bibfnamefont {C.~J.}\ \bibnamefont
  {Hughes}},\ }\bibfield  {title} {\enquote {\bibinfo {title} {High-frequency
  capillary waves on the clean surface of water},}\ }\href {\doibase
  10.1021/la00059a002} {\bibfield  {journal} {\bibinfo  {journal} {Langmuir}\
  }\textbf {\bibinfo {volume} {7}},\ \bibinfo {pages} {2419--2421} (\bibinfo
  {year} {1991})}\BibitemShut {NoStop}%
\bibitem [{\citenamefont {F{\'a}bi{\'a}n}\ \emph {et~al.}(2017)\citenamefont
  {F{\'a}bi{\'a}n}, \citenamefont {Sega}, \citenamefont {Horvai},\ and\
  \citenamefont {Jedlovszky}}]{fabian_single_2017}%
  \BibitemOpen
  \bibfield  {author} {\bibinfo {author} {\bibfnamefont {B.}~\bibnamefont
  {F{\'a}bi{\'a}n}}, \bibinfo {author} {\bibfnamefont {M.}~\bibnamefont
  {Sega}}, \bibinfo {author} {\bibfnamefont {G.}~\bibnamefont {Horvai}}, \ and\
  \bibinfo {author} {\bibfnamefont {P.}~\bibnamefont {Jedlovszky}},\ }\bibfield
   {title} {\enquote {\bibinfo {title} {Single {{Particle Dynamics}} at the
  {{Intrinsic Surface}} of {{Various Apolar}}, {{Aprotic Dipolar}}, and
  {{Hydrogen Bonding Liquids As Seen}} from {{Computer Simulations}}},}\ }\href
  {\doibase 10.1021/acs.jpcb.7b02220} {\bibfield  {journal} {\bibinfo
  {journal} {The Journal of Physical Chemistry B}\ }\textbf {\bibinfo {volume}
  {121}},\ \bibinfo {pages} {5582--5594} (\bibinfo {year} {2017})}\BibitemShut
  {NoStop}%
\bibitem [{\citenamefont {{Ruiz-Lopez}}\ \emph {et~al.}(2020)\citenamefont
  {{Ruiz-Lopez}}, \citenamefont {Francisco}, \citenamefont {{Martins-Costa}},\
  and\ \citenamefont {Anglada}}]{ruiz-lopez_molecular_2020}%
  \BibitemOpen
  \bibfield  {author} {\bibinfo {author} {\bibfnamefont {M.~F.}\ \bibnamefont
  {{Ruiz-Lopez}}}, \bibinfo {author} {\bibfnamefont {J.~S.}\ \bibnamefont
  {Francisco}}, \bibinfo {author} {\bibfnamefont {M.~T.}\ \bibnamefont
  {{Martins-Costa}}}, \ and\ \bibinfo {author} {\bibfnamefont {J.~M.}\
  \bibnamefont {Anglada}},\ }\bibfield  {title} {\enquote {\bibinfo {title}
  {Molecular reactions at aqueous interfaces},}\ }\href@noop {} {\bibfield
  {journal} {\bibinfo  {journal} {Nature Reviews Chemistry}\ }\textbf {\bibinfo
  {volume} {4}},\ \bibinfo {pages} {459--475} (\bibinfo {year}
  {2020})}\BibitemShut {NoStop}%
\bibitem [{\citenamefont {Narayan}\ \emph {et~al.}(2005)\citenamefont
  {Narayan}, \citenamefont {Muldoon}, \citenamefont {Finn}, \citenamefont
  {Fokin}, \citenamefont {Kolb},\ and\ \citenamefont
  {Sharpless}}]{narayan_water_2005}%
  \BibitemOpen
  \bibfield  {author} {\bibinfo {author} {\bibfnamefont {S.}~\bibnamefont
  {Narayan}}, \bibinfo {author} {\bibfnamefont {J.}~\bibnamefont {Muldoon}},
  \bibinfo {author} {\bibfnamefont {M.~G.}\ \bibnamefont {Finn}}, \bibinfo
  {author} {\bibfnamefont {V.~V.}\ \bibnamefont {Fokin}}, \bibinfo {author}
  {\bibfnamefont {H.~C.}\ \bibnamefont {Kolb}}, \ and\ \bibinfo {author}
  {\bibfnamefont {K.~B.}\ \bibnamefont {Sharpless}},\ }\bibfield  {title}
  {\enquote {\bibinfo {title} {``{{On}} water'': Unique reactivity of organic
  compounds in aqueous suspension},}\ }\href@noop {} {\bibfield  {journal}
  {\bibinfo  {journal} {Angewandte Chemie International Edition}\ }\textbf
  {\bibinfo {volume} {44}},\ \bibinfo {pages} {3275--3279} (\bibinfo {year}
  {2005})}\BibitemShut {NoStop}%
\bibitem [{\citenamefont {Jung}\ and\ \citenamefont
  {Marcus}(2007)}]{jung_theory_2007}%
  \BibitemOpen
  \bibfield  {author} {\bibinfo {author} {\bibfnamefont {Y.}~\bibnamefont
  {Jung}}\ and\ \bibinfo {author} {\bibfnamefont {R.~A.}\ \bibnamefont
  {Marcus}},\ }\bibfield  {title} {\enquote {\bibinfo {title} {On the theory of
  organic catalysis ``on water''},}\ }\href {\doibase 10.1021/ja068120f}
  {\bibfield  {journal} {\bibinfo  {journal} {Journal of the American Chemical
  Society}\ }\textbf {\bibinfo {volume} {129}},\ \bibinfo {pages} {5492--5502}
  (\bibinfo {year} {2007})}\BibitemShut {NoStop}%
\bibitem [{\citenamefont {Yan}, \citenamefont {Bain},\ and\ \citenamefont
  {Cooks}(2016)}]{yan_organic_2016}%
  \BibitemOpen
  \bibfield  {author} {\bibinfo {author} {\bibfnamefont {X.}~\bibnamefont
  {Yan}}, \bibinfo {author} {\bibfnamefont {R.~M.}\ \bibnamefont {Bain}}, \
  and\ \bibinfo {author} {\bibfnamefont {R.~G.}\ \bibnamefont {Cooks}},\
  }\bibfield  {title} {\enquote {\bibinfo {title} {Organic reactions in
  microdroplets: {{Reaction}} acceleration revealed by mass spectrometry},}\
  }\href@noop {} {\bibfield  {journal} {\bibinfo  {journal} {Angewandte Chemie
  International Edition}\ }\textbf {\bibinfo {volume} {55}},\ \bibinfo {pages}
  {12960--12972} (\bibinfo {year} {2016})}\BibitemShut {NoStop}%
\bibitem [{\citenamefont {Boda}\ \emph {et~al.}(1996)\citenamefont {Boda},
  \citenamefont {Luk{\'a}cs}, \citenamefont {Liszi},\ and\ \citenamefont
  {Szalai}}]{boda_isochoric-_1996}%
  \BibitemOpen
  \bibfield  {author} {\bibinfo {author} {\bibfnamefont {D.}~\bibnamefont
  {Boda}}, \bibinfo {author} {\bibfnamefont {T.}~\bibnamefont {Luk{\'a}cs}},
  \bibinfo {author} {\bibfnamefont {J.}~\bibnamefont {Liszi}}, \ and\ \bibinfo
  {author} {\bibfnamefont {I.}~\bibnamefont {Szalai}},\ }\bibfield  {title}
  {\enquote {\bibinfo {title} {The isochoric-, isobaric-and saturation-heat
  capacities of the {{Lennard-Jones}} fluid from equations of state and {{Monte
  Carlo}} simulations},}\ }\href {\doibase 10.1016/0378-3812(96)02998-6}
  {\bibfield  {journal} {\bibinfo  {journal} {Fluid phase equilibria}\ }\textbf
  {\bibinfo {volume} {119}},\ \bibinfo {pages} {1--16} (\bibinfo {year}
  {1996})}\BibitemShut {NoStop}%
\bibitem [{\citenamefont {Essmann}\ \emph {et~al.}(1995)\citenamefont
  {Essmann}, \citenamefont {Perera}, \citenamefont {Berkowitz}, \citenamefont
  {Darden}, \citenamefont {Lee},\ and\ \citenamefont
  {Pedersen}}]{essmann_smooth_1995}%
  \BibitemOpen
  \bibfield  {author} {\bibinfo {author} {\bibfnamefont {U.}~\bibnamefont
  {Essmann}}, \bibinfo {author} {\bibfnamefont {L.}~\bibnamefont {Perera}},
  \bibinfo {author} {\bibfnamefont {M.~L.}\ \bibnamefont {Berkowitz}}, \bibinfo
  {author} {\bibfnamefont {T.}~\bibnamefont {Darden}}, \bibinfo {author}
  {\bibfnamefont {H.}~\bibnamefont {Lee}}, \ and\ \bibinfo {author}
  {\bibfnamefont {L.~G.}\ \bibnamefont {Pedersen}},\ }\bibfield  {title}
  {\enquote {\bibinfo {title} {A smooth particle mesh {{Ewald}} method},}\
  }\href {\doibase 10.1063/1.470117} {\bibfield  {journal} {\bibinfo  {journal}
  {J. Chem. Phys.}\ }\textbf {\bibinfo {volume} {103}},\ \bibinfo {pages}
  {8577--8593} (\bibinfo {year} {1995})}\BibitemShut {NoStop}%
\bibitem [{\citenamefont {Nos{\'e}}(1984)}]{nose_molecular_1984}%
  \BibitemOpen
  \bibfield  {author} {\bibinfo {author} {\bibfnamefont {S.}~\bibnamefont
  {Nos{\'e}}},\ }\bibfield  {title} {\enquote {\bibinfo {title} {A molecular
  dynamics method for simulations in the canonical ensemble},}\ }\href@noop {}
  {\bibfield  {journal} {\bibinfo  {journal} {Mol. Phys.}\ }\textbf {\bibinfo
  {volume} {52}},\ \bibinfo {pages} {255--268} (\bibinfo {year}
  {1984})}\BibitemShut {NoStop}%
\bibitem [{\citenamefont {Hoover}(1985)}]{hoover_canonical_1985}%
  \BibitemOpen
  \bibfield  {author} {\bibinfo {author} {\bibfnamefont {W.~G.}\ \bibnamefont
  {Hoover}},\ }\bibfield  {title} {\enquote {\bibinfo {title} {Canonical
  dynamics: {{Equilibrium}} phase-space distributions},}\ }\href {\doibase
  10.1103/PhysRevA.31.1695} {\bibfield  {journal} {\bibinfo  {journal} {Phys.
  Rev. A}\ }\textbf {\bibinfo {volume} {31}},\ \bibinfo {pages} {1695}
  (\bibinfo {year} {1985})}\BibitemShut {NoStop}%
\bibitem [{\citenamefont {Michaud-Agrawal}\ \emph {et~al.}(2011)\citenamefont
  {Michaud-Agrawal}, \citenamefont {Denning}, \citenamefont {Woolf},\ and\
  \citenamefont {Beckstein}}]{michaudagrawal_mdanalysis_2011}%
  \BibitemOpen
  \bibfield  {author} {\bibinfo {author} {\bibfnamefont {N.}~\bibnamefont
  {Michaud-Agrawal}}, \bibinfo {author} {\bibfnamefont {E.~J.}\ \bibnamefont
  {Denning}}, \bibinfo {author} {\bibfnamefont {T.~B.}\ \bibnamefont {Woolf}},
  \ and\ \bibinfo {author} {\bibfnamefont {O.}~\bibnamefont {Beckstein}},\
  }\bibfield  {title} {\enquote {\bibinfo {title} {{{MDAnalysis}}: A toolkit
  for the analysis of molecular dynamics simulations},}\ }\href@noop {}
  {\bibfield  {journal} {\bibinfo  {journal} {J. Comput. Chem.}\ }\textbf
  {\bibinfo {volume} {32}},\ \bibinfo {pages} {2319--2327} (\bibinfo {year}
  {2011})}\BibitemShut {NoStop}%
\bibitem [{\citenamefont {Sega}\ and\ \citenamefont
  {Hantal}(2017)}]{sega_phase_2017}%
  \BibitemOpen
  \bibfield  {author} {\bibinfo {author} {\bibfnamefont {M.}~\bibnamefont
  {Sega}}\ and\ \bibinfo {author} {\bibfnamefont {G.}~\bibnamefont {Hantal}},\
  }\bibfield  {title} {\enquote {\bibinfo {title} {Phase and interface
  determination in computer simulations of liquid mixtures with high partial
  miscibility},}\ }\href {\doibase 10.1039/C7CP02918G} {\bibfield  {journal}
  {\bibinfo  {journal} {Phys. Chem. Chem. Phys.}\ }\textbf {\bibinfo {volume}
  {19}},\ \bibinfo {pages} {18968--18974} (\bibinfo {year} {2017})}\BibitemShut
  {NoStop}%
\bibitem [{\citenamefont {Rahman}(1968)}]{rahman_current_1968}%
  \BibitemOpen
  \bibfield  {author} {\bibinfo {author} {\bibfnamefont {A.}~\bibnamefont
  {Rahman}},\ }\bibfield  {title} {\enquote {\bibinfo {title} {Current
  fluctuations in classical liquids},}\ }in\ \href@noop {} {\emph {\bibinfo
  {booktitle} {Neutron {{Inelastic Scattering Vol}}. {{I}}. {{Proceedings}} of
  a {{Symposium}} on {{Neutron Inelastic Scattering}}}}}\ (\bibinfo {year}
  {1968})\BibitemShut {NoStop}%
\bibitem [{\citenamefont {Boon}\ and\ \citenamefont
  {Yip}(1980)}]{boon_molecular_1980}%
  \BibitemOpen
  \bibfield  {author} {\bibinfo {author} {\bibfnamefont {J.~P.}\ \bibnamefont
  {Boon}}\ and\ \bibinfo {author} {\bibfnamefont {S.}~\bibnamefont {Yip}},\
  }\href@noop {} {\emph {\bibinfo {title} {Molecular {{Hydrodynamics}}}}}\
  (\bibinfo  {publisher} {{McGraw-Hill Inc.}},\ \bibinfo {year}
  {1980})\BibitemShut {NoStop}%
\bibitem [{\citenamefont {Levenberg}(1944)}]{levenberg_method_1944}%
  \BibitemOpen
  \bibfield  {author} {\bibinfo {author} {\bibfnamefont {K.}~\bibnamefont
  {Levenberg}},\ }\bibfield  {title} {\enquote {\bibinfo {title} {A method for
  the solution of certain non-linear problems in least squares},}\ }\href@noop
  {} {\bibfield  {journal} {\bibinfo  {journal} {Quarterly of applied
  mathematics}\ }\textbf {\bibinfo {volume} {2}},\ \bibinfo {pages} {164--168}
  (\bibinfo {year} {1944})}\BibitemShut {NoStop}%
\bibitem [{\citenamefont {Marquardt}(1963)}]{marquardt_algorithm_1963}%
  \BibitemOpen
  \bibfield  {author} {\bibinfo {author} {\bibfnamefont {D.~W.}\ \bibnamefont
  {Marquardt}},\ }\bibfield  {title} {\enquote {\bibinfo {title} {An algorithm
  for least-squares estimation of nonlinear parameters},}\ }\href@noop {}
  {\bibfield  {journal} {\bibinfo  {journal} {Journal of the society for
  Industrial and Applied Mathematics}\ }\textbf {\bibinfo {volume} {11}},\
  \bibinfo {pages} {431--441} (\bibinfo {year} {1963})}\BibitemShut {NoStop}%
\bibitem [{\citenamefont {Falk}\ and\ \citenamefont
  {Mecke}(2011)}]{falk_capillary_2011}%
  \BibitemOpen
  \bibfield  {author} {\bibinfo {author} {\bibfnamefont {K.}~\bibnamefont
  {Falk}}\ and\ \bibinfo {author} {\bibfnamefont {K.}~\bibnamefont {Mecke}},\
  }\bibfield  {title} {\enquote {\bibinfo {title} {Capillary waves of
  compressible fluids},}\ }\href {\doibase 10.1088/0953-8984/23/18/184103}
  {\bibfield  {journal} {\bibinfo  {journal} {J. Phys.: Condens. Matter}\
  }\textbf {\bibinfo {volume} {23}},\ \bibinfo {pages} {184103} (\bibinfo
  {year} {2011})}\BibitemShut {NoStop}%
\bibitem [{\citenamefont {Hansen}\ and\ \citenamefont
  {McDonald}(2013)}]{hansen2013theory}%
  \BibitemOpen
  \bibfield  {author} {\bibinfo {author} {\bibfnamefont {J.~P.}\ \bibnamefont
  {Hansen}}\ and\ \bibinfo {author} {\bibfnamefont {I.~R.}\ \bibnamefont
  {McDonald}},\ }\href@noop {} {\emph {\bibinfo {title} {Theory of Simple
  Liquids: With Applications to Soft Matter}}}\ (\bibinfo  {publisher}
  {{Elsevier Science}},\ \bibinfo {year} {2013})\BibitemShut {NoStop}%
\bibitem [{\citenamefont {Savitzky}\ and\ \citenamefont
  {Golay}(1964)}]{savitzky_smoothing_1964}%
  \BibitemOpen
  \bibfield  {author} {\bibinfo {author} {\bibfnamefont {A.}~\bibnamefont
  {Savitzky}}\ and\ \bibinfo {author} {\bibfnamefont {M.~J.}\ \bibnamefont
  {Golay}},\ }\bibfield  {title} {\enquote {\bibinfo {title} {Smoothing and
  differentiation of data by simplified least squares procedures.}}\
  }\href@noop {} {\bibfield  {journal} {\bibinfo  {journal} {Analytical
  chemistry}\ }\textbf {\bibinfo {volume} {36}},\ \bibinfo {pages} {1627--1639}
  (\bibinfo {year} {1964})}\BibitemShut {NoStop}%
\bibitem [{\citenamefont {Loudon}(1980)}]{loudon_theory_1980}%
  \BibitemOpen
  \bibfield  {author} {\bibinfo {author} {\bibfnamefont {R.}~\bibnamefont
  {Loudon}},\ }\bibfield  {title} {\enquote {\bibinfo {title} {Theory of
  thermally induced surface fluctuations on simple fluids},}\ }\href@noop {}
  {\bibfield  {journal} {\bibinfo  {journal} {Proceedings of the Royal Society
  of London. A. Mathematical and Physical Sciences}\ }\textbf {\bibinfo
  {volume} {372}},\ \bibinfo {pages} {275--295} (\bibinfo {year}
  {1980})}\BibitemShut {NoStop}%
\bibitem [{\citenamefont {Harden}, \citenamefont {Pleiner},\ and\ \citenamefont
  {Pincus}(1991)}]{harden_hydrodynamic_1991}%
  \BibitemOpen
  \bibfield  {author} {\bibinfo {author} {\bibfnamefont {J.~L.}\ \bibnamefont
  {Harden}}, \bibinfo {author} {\bibfnamefont {H.}~\bibnamefont {Pleiner}}, \
  and\ \bibinfo {author} {\bibfnamefont {P.~A.}\ \bibnamefont {Pincus}},\
  }\bibfield  {title} {\enquote {\bibinfo {title} {Hydrodynamic surface modes
  on concentrated polymer solutions and gels},}\ }\href {\doibase
  10.1063/1.460525} {\bibfield  {journal} {\bibinfo  {journal} {The Journal of
  chemical physics}\ }\textbf {\bibinfo {volume} {94}},\ \bibinfo {pages}
  {5208--5221} (\bibinfo {year} {1991})}\BibitemShut {NoStop}%
\bibitem [{\citenamefont {Mora}\ and\ \citenamefont
  {Daillant}(2002)}]{mora_height_2002}%
  \BibitemOpen
  \bibfield  {author} {\bibinfo {author} {\bibfnamefont {S.}~\bibnamefont
  {Mora}}\ and\ \bibinfo {author} {\bibfnamefont {J.}~\bibnamefont
  {Daillant}},\ }\bibfield  {title} {\enquote {\bibinfo {title} {Height and
  density correlations at liquid surfaces; application to {{X-ray}}
  scattering},}\ }\href@noop {} {\bibfield  {journal} {\bibinfo  {journal} {The
  European Physical Journal B-Condensed Matter and Complex Systems}\ }\textbf
  {\bibinfo {volume} {27}},\ \bibinfo {pages} {417--428} (\bibinfo {year}
  {2002})}\BibitemShut {NoStop}%
\bibitem [{\citenamefont {Chung}\ and\ \citenamefont
  {Yip}(1969)}]{chung_generalized_1969}%
  \BibitemOpen
  \bibfield  {author} {\bibinfo {author} {\bibfnamefont {C.-H.}\ \bibnamefont
  {Chung}}\ and\ \bibinfo {author} {\bibfnamefont {S.}~\bibnamefont {Yip}},\
  }\bibfield  {title} {\enquote {\bibinfo {title} {Generalized hydrodynamics
  and time correlation functions},}\ }\href {\doibase 10.1103/PhysRev.182.323}
  {\bibfield  {journal} {\bibinfo  {journal} {Physical Review}\ }\textbf
  {\bibinfo {volume} {182}},\ \bibinfo {pages} {323} (\bibinfo {year}
  {1969})}\BibitemShut {NoStop}%
\bibitem [{\citenamefont {Gaskell}\ \emph {et~al.}(1987)\citenamefont
  {Gaskell}, \citenamefont {Balucani}, \citenamefont {Gori},\ and\
  \citenamefont {Vallauri}}]{gaskell_wavevector-dependent_1987}%
  \BibitemOpen
  \bibfield  {author} {\bibinfo {author} {\bibfnamefont {T.}~\bibnamefont
  {Gaskell}}, \bibinfo {author} {\bibfnamefont {U.}~\bibnamefont {Balucani}},
  \bibinfo {author} {\bibfnamefont {M.}~\bibnamefont {Gori}}, \ and\ \bibinfo
  {author} {\bibfnamefont {R.}~\bibnamefont {Vallauri}},\ }\bibfield  {title}
  {\enquote {\bibinfo {title} {Wavevector-dependent shear viscosity in
  {{Lennard-Jones}} liquids},}\ }\href {\doibase 10.1088/0031-8949/35/1/007}
  {\bibfield  {journal} {\bibinfo  {journal} {Physica scripta}\ }\textbf
  {\bibinfo {volume} {35}},\ \bibinfo {pages} {37} (\bibinfo {year}
  {1987})}\BibitemShut {NoStop}%
\bibitem [{\citenamefont {Mecke}\ and\ \citenamefont
  {Dietrich}(1999)}]{mecke_effective_1999}%
  \BibitemOpen
  \bibfield  {author} {\bibinfo {author} {\bibfnamefont {K.~R.}\ \bibnamefont
  {Mecke}}\ and\ \bibinfo {author} {\bibfnamefont {S.}~\bibnamefont
  {Dietrich}},\ }\bibfield  {title} {\enquote {\bibinfo {title} {Effective
  {{Hamiltonian}} for liquid-vapor interfaces},}\ }\href {\doibase
  10.1103/PhysRevE.59.6766} {\bibfield  {journal} {\bibinfo  {journal} {Phys.
  Rev. E}\ }\textbf {\bibinfo {volume} {59}},\ \bibinfo {pages} {6766--6784}
  (\bibinfo {year} {1999})}\BibitemShut {NoStop}%
\bibitem [{\citenamefont {H{\"o}fling}\ and\ \citenamefont
  {Dietrich}(2015)}]{hofling_enhanced_2015}%
  \BibitemOpen
  \bibfield  {author} {\bibinfo {author} {\bibfnamefont {F.}~\bibnamefont
  {H{\"o}fling}}\ and\ \bibinfo {author} {\bibfnamefont {S.}~\bibnamefont
  {Dietrich}},\ }\bibfield  {title} {\enquote {\bibinfo {title} {Enhanced
  wavelength-dependent surface tension of liquid-vapour interfaces},}\ }\href
  {\doibase 10.1209/0295-5075/109/46002} {\bibfield  {journal} {\bibinfo
  {journal} {Europhys. Lett.}\ }\textbf {\bibinfo {volume} {109}},\ \bibinfo
  {pages} {46002} (\bibinfo {year} {2015})}\BibitemShut {NoStop}%
\bibitem [{\citenamefont {{Delgado-Buscalioni}}, \citenamefont {Chacon},\ and\
  \citenamefont {Tarazona}(2008)}]{delgado-buscalioni_hydrodynamics_2008}%
  \BibitemOpen
  \bibfield  {author} {\bibinfo {author} {\bibfnamefont {R.}~\bibnamefont
  {{Delgado-Buscalioni}}}, \bibinfo {author} {\bibfnamefont {E.}~\bibnamefont
  {Chacon}}, \ and\ \bibinfo {author} {\bibfnamefont {P.}~\bibnamefont
  {Tarazona}},\ }\bibfield  {title} {\enquote {\bibinfo {title} {Hydrodynamics
  of {{Nanoscopic Capillary Waves}}},}\ }\href {\doibase
  10.1103/PhysRevLett.101.106102} {\bibfield  {journal} {\bibinfo  {journal}
  {Phys. Rev. Lett.}\ }\textbf {\bibinfo {volume} {101}},\ \bibinfo {pages}
  {106102} (\bibinfo {year} {2008})}\BibitemShut {NoStop}%
\bibitem [{\citenamefont {{Hern{\'a}ndez-Mu{\~n}oz}}, \citenamefont
  {Tarazona},\ and\ \citenamefont
  {Chacon}(2022)}]{hernandez-munoz_layering_2022}%
  \BibitemOpen
  \bibfield  {author} {\bibinfo {author} {\bibfnamefont {J.}~\bibnamefont
  {{Hern{\'a}ndez-Mu{\~n}oz}}}, \bibinfo {author} {\bibfnamefont
  {P.}~\bibnamefont {Tarazona}}, \ and\ \bibinfo {author} {\bibfnamefont
  {E.}~\bibnamefont {Chacon}},\ }\bibfield  {title} {\enquote {\bibinfo {title}
  {Layering and capillary waves in the structure factor of liquid surfaces},}\
  }\href {\doibase ; https://doi.org/10.1063/5.0118252} {\bibfield  {journal}
  {\bibinfo  {journal} {J. Chem. Phys.}\ }\textbf {\bibinfo {volume} {157}},\
  \bibinfo {pages} {154703} (\bibinfo {year} {2022})}\BibitemShut {NoStop}%
\bibitem [{\citenamefont {Hansen}\ \emph {et~al.}(2007)\citenamefont {Hansen},
  \citenamefont {Daivis}, \citenamefont {Travis},\ and\ \citenamefont
  {Todd}}]{hansen2007parameterization}%
  \BibitemOpen
  \bibfield  {author} {\bibinfo {author} {\bibfnamefont {J.}~\bibnamefont
  {Hansen}}, \bibinfo {author} {\bibfnamefont {P.~J.}\ \bibnamefont {Daivis}},
  \bibinfo {author} {\bibfnamefont {K.~P.}\ \bibnamefont {Travis}}, \ and\
  \bibinfo {author} {\bibfnamefont {B.}~\bibnamefont {Todd}},\ }\bibfield
  {title} {\enquote {\bibinfo {title} {Parameterization of the nonlocal
  viscosity kernel for an atomic fluid},}\ }\href@noop {} {\bibfield  {journal}
  {\bibinfo  {journal} {Physical Review E}\ }\textbf {\bibinfo {volume} {76}},\
  \bibinfo {pages} {041121} (\bibinfo {year} {2007})}\BibitemShut {NoStop}%
\bibitem [{\citenamefont {Duque-Zumajo}, \citenamefont {De~La~Torre},\ and\
  \citenamefont {Espa{\~n}ol}(2020)}]{duque2020non}%
  \BibitemOpen
  \bibfield  {author} {\bibinfo {author} {\bibfnamefont {D.}~\bibnamefont
  {Duque-Zumajo}}, \bibinfo {author} {\bibfnamefont {J.}~\bibnamefont
  {De~La~Torre}}, \ and\ \bibinfo {author} {\bibfnamefont {P.}~\bibnamefont
  {Espa{\~n}ol}},\ }\bibfield  {title} {\enquote {\bibinfo {title} {Non-local
  viscosity from the green--kubo formula},}\ }\href@noop {} {\bibfield
  {journal} {\bibinfo  {journal} {The Journal of Chemical Physics}\ }\textbf
  {\bibinfo {volume} {152}},\ \bibinfo {pages} {174108} (\bibinfo {year}
  {2020})}\BibitemShut {NoStop}%
\bibitem [{\citenamefont {Malgaretti}\ \emph {et~al.}(2023)\citenamefont
  {Malgaretti}, \citenamefont {Bafile}, \citenamefont {Vallauri}, \citenamefont
  {Jedlovszki},\ and\ \citenamefont {Sega}}]{zenodo7416368}%
  \BibitemOpen
  \bibfield  {author} {\bibinfo {author} {\bibfnamefont {P.}~\bibnamefont
  {Malgaretti}}, \bibinfo {author} {\bibfnamefont {U.}~\bibnamefont {Bafile}},
  \bibinfo {author} {\bibfnamefont {R.}~\bibnamefont {Vallauri}}, \bibinfo
  {author} {\bibfnamefont {P.}~\bibnamefont {Jedlovszki}}, \ and\ \bibinfo
  {author} {\bibfnamefont {M.}~\bibnamefont {Sega}},\ }\href {\doibase
  10.5281/zenodo.7416368} {\enquote {\bibinfo {title} {Dataset for ``surface
  viscosity in simple liquids''},}\ }\bibinfo {howpublished} {zenodo} (\bibinfo
  {year} {2023}),\ \bibinfo {note}
  {{https://www.zenodo.org/record/7416368}}\BibitemShut {NoStop}%
\bibitem [{\citenamefont {P{\'a}rtay}\ \emph {et~al.}(2008)\citenamefont
  {P{\'a}rtay}, \citenamefont {Hantal}, \citenamefont {Jedlovszky},
  \citenamefont {Vincze},\ and\ \citenamefont {Horvai}}]{partay_new_2008}%
  \BibitemOpen
  \bibfield  {author} {\bibinfo {author} {\bibfnamefont {L.~B.}\ \bibnamefont
  {P{\'a}rtay}}, \bibinfo {author} {\bibfnamefont {G.}~\bibnamefont {Hantal}},
  \bibinfo {author} {\bibfnamefont {P.}~\bibnamefont {Jedlovszky}}, \bibinfo
  {author} {\bibfnamefont {{\'A}.}~\bibnamefont {Vincze}}, \ and\ \bibinfo
  {author} {\bibfnamefont {G.}~\bibnamefont {Horvai}},\ }\bibfield  {title}
  {\enquote {\bibinfo {title} {A new method for determining the interfacial
  molecules and characterizing the surface roughness in computer simulations.
  {{Application}} to the liquid\textendash vapor interface of water},}\
  }\href@noop {} {\bibfield  {journal} {\bibinfo  {journal} {J. Comput. Chem.}\
  }\textbf {\bibinfo {volume} {29}},\ \bibinfo {pages} {945--956} (\bibinfo
  {year} {2008})}\BibitemShut {NoStop}%
\bibitem [{\citenamefont {Sega}\ \emph {et~al.}(2018)\citenamefont {Sega},
  \citenamefont {Hantal}, \citenamefont {F{\'a}bi{\'a}n},\ and\ \citenamefont
  {Jedlovszky}}]{sega_pytim_2018}%
  \BibitemOpen
  \bibfield  {author} {\bibinfo {author} {\bibfnamefont {M.}~\bibnamefont
  {Sega}}, \bibinfo {author} {\bibfnamefont {G.}~\bibnamefont {Hantal}},
  \bibinfo {author} {\bibfnamefont {B.}~\bibnamefont {F{\'a}bi{\'a}n}}, \ and\
  \bibinfo {author} {\bibfnamefont {P.}~\bibnamefont {Jedlovszky}},\ }\bibfield
   {title} {\enquote {\bibinfo {title} {Pytim: {{A}} python package for the
  interfacial analysis of molecular simulations},}\ }\href {\doibase
  10.1002/jcc.25384} {\bibfield  {journal} {\bibinfo  {journal} {J. Comput.
  Chem.}\ }\textbf {\bibinfo {volume} {39}},\ \bibinfo {pages} {2118--2125}
  (\bibinfo {year} {2018})}\BibitemShut {NoStop}%
\bibitem [{\citenamefont {Giri}\ \emph {et~al.}(2022)\citenamefont {Giri},
  \citenamefont {Malgaretti}, \citenamefont {Peschka},\ and\ \citenamefont
  {Sega}}]{giri_resolving_2022}%
  \BibitemOpen
  \bibfield  {author} {\bibinfo {author} {\bibfnamefont {A.~K.}\ \bibnamefont
  {Giri}}, \bibinfo {author} {\bibfnamefont {P.}~\bibnamefont {Malgaretti}},
  \bibinfo {author} {\bibfnamefont {D.}~\bibnamefont {Peschka}}, \ and\
  \bibinfo {author} {\bibfnamefont {M.}~\bibnamefont {Sega}},\ }\bibfield
  {title} {\enquote {\bibinfo {title} {Resolving the microscopic hydrodynamics
  at the moving contact line},}\ }\href {\doibase
  10.1103/PhysRevFluids.7.L102001} {\bibfield  {journal} {\bibinfo  {journal}
  {Phys. Rev. Fluids}\ }\textbf {\bibinfo {volume} {7}},\ \bibinfo {pages}
  {L102001} (\bibinfo {year} {2022})}\BibitemShut {NoStop}%
\bibitem [{\citenamefont {Chac{\'o}n}\ and\ \citenamefont
  {Tarazona}(2003)}]{chacon_intrinsic_2003}%
  \BibitemOpen
  \bibfield  {author} {\bibinfo {author} {\bibfnamefont {E.}~\bibnamefont
  {Chac{\'o}n}}\ and\ \bibinfo {author} {\bibfnamefont {P.}~\bibnamefont
  {Tarazona}},\ }\bibfield  {title} {\enquote {\bibinfo {title} {Intrinsic
  profiles beyond the capillary wave theory: {{A Monte Carlo}} study},}\ }\href
  {\doibase 10.1103/PhysRevLett.91.166103} {\bibfield  {journal} {\bibinfo
  {journal} {Phys. Rev. Lett.}\ }\textbf {\bibinfo {volume} {91}},\ \bibinfo
  {pages} {166103} (\bibinfo {year} {2003})}\BibitemShut {NoStop}%
\bibitem [{\citenamefont {Sega}, \citenamefont {F{\'a}bi{\'a}n},\ and\
  \citenamefont {Jedlovszky}(2015)}]{sega_layer-by-layer_2015}%
  \BibitemOpen
  \bibfield  {author} {\bibinfo {author} {\bibfnamefont {M.}~\bibnamefont
  {Sega}}, \bibinfo {author} {\bibfnamefont {B.}~\bibnamefont
  {F{\'a}bi{\'a}n}}, \ and\ \bibinfo {author} {\bibfnamefont {P.}~\bibnamefont
  {Jedlovszky}},\ }\bibfield  {title} {\enquote {\bibinfo {title}
  {Layer-by-layer and intrinsic analysis of molecular and thermodynamic
  properties across soft interfaces},}\ }\href {\doibase 10.1063/1.4931180}
  {\bibfield  {journal} {\bibinfo  {journal} {J. Chem. Phys.}\ }\textbf
  {\bibinfo {volume} {143}},\ \bibinfo {pages} {114709} (\bibinfo {year}
  {2015})}\BibitemShut {NoStop}%
\bibitem [{\citenamefont {Sendner}\ \emph {et~al.}(2009)\citenamefont
  {Sendner}, \citenamefont {Horinek}, \citenamefont {Bocquet},\ and\
  \citenamefont {Netz}}]{sendner_interfacial_2009}%
  \BibitemOpen
  \bibfield  {author} {\bibinfo {author} {\bibfnamefont {C.}~\bibnamefont
  {Sendner}}, \bibinfo {author} {\bibfnamefont {D.}~\bibnamefont {Horinek}},
  \bibinfo {author} {\bibfnamefont {L.}~\bibnamefont {Bocquet}}, \ and\
  \bibinfo {author} {\bibfnamefont {R.~R.}\ \bibnamefont {Netz}},\ }\bibfield
  {title} {\enquote {\bibinfo {title} {Interfacial {{Water}} at {{Hydrophobic}}
  and {{Hydrophilic Surfaces}}: {{Slip}}, {{Viscosity}}, and {{Diffusion}}},}\
  }\href {\doibase 10.1021/la901314b} {\bibfield  {journal} {\bibinfo
  {journal} {Langmuir}\ }\textbf {\bibinfo {volume} {25}},\ \bibinfo {pages}
  {10768--10781} (\bibinfo {year} {2009})}\BibitemShut {NoStop}%
\bibitem [{\citenamefont {Thompson}\ and\ \citenamefont
  {Robbins}(1990)}]{thompson1990shear}%
  \BibitemOpen
  \bibfield  {author} {\bibinfo {author} {\bibfnamefont {P.~A.}\ \bibnamefont
  {Thompson}}\ and\ \bibinfo {author} {\bibfnamefont {M.~O.}\ \bibnamefont
  {Robbins}},\ }\bibfield  {title} {\enquote {\bibinfo {title} {Shear flow near
  solids: Epitaxial order and flow boundary conditions},}\ }\href@noop {}
  {\bibfield  {journal} {\bibinfo  {journal} {Physical review A}\ }\textbf
  {\bibinfo {volume} {41}},\ \bibinfo {pages} {6830} (\bibinfo {year}
  {1990})}\BibitemShut {NoStop}%
\bibitem [{\citenamefont {Hansen}, \citenamefont {Todd},\ and\ \citenamefont
  {Daivis}(2011)}]{hansen2011prediction}%
  \BibitemOpen
  \bibfield  {author} {\bibinfo {author} {\bibfnamefont {J.~S.}\ \bibnamefont
  {Hansen}}, \bibinfo {author} {\bibfnamefont {B.}~\bibnamefont {Todd}}, \ and\
  \bibinfo {author} {\bibfnamefont {P.~J.}\ \bibnamefont {Daivis}},\ }\bibfield
   {title} {\enquote {\bibinfo {title} {Prediction of fluid velocity slip at
  solid surfaces},}\ }\href@noop {} {\bibfield  {journal} {\bibinfo  {journal}
  {Physical Review E}\ }\textbf {\bibinfo {volume} {84}},\ \bibinfo {pages}
  {016313} (\bibinfo {year} {2011})}\BibitemShut {NoStop}%
\bibitem [{\citenamefont {Ho}\ \emph {et~al.}(2011)\citenamefont {Ho},
  \citenamefont {Papavassiliou}, \citenamefont {Lee},\ and\ \citenamefont
  {Striolo}}]{ho2011liquid}%
  \BibitemOpen
  \bibfield  {author} {\bibinfo {author} {\bibfnamefont {T.~A.}\ \bibnamefont
  {Ho}}, \bibinfo {author} {\bibfnamefont {D.~V.}\ \bibnamefont
  {Papavassiliou}}, \bibinfo {author} {\bibfnamefont {L.~L.}\ \bibnamefont
  {Lee}}, \ and\ \bibinfo {author} {\bibfnamefont {A.}~\bibnamefont
  {Striolo}},\ }\bibfield  {title} {\enquote {\bibinfo {title} {Liquid water
  can slip on a hydrophilic surface},}\ }\href@noop {} {\bibfield  {journal}
  {\bibinfo  {journal} {Proceedings of the National Academy of Sciences}\
  }\textbf {\bibinfo {volume} {108}},\ \bibinfo {pages} {16170--16175}
  (\bibinfo {year} {2011})}\BibitemShut {NoStop}%
\bibitem [{\citenamefont {Sega}\ \emph {et~al.}(2013)\citenamefont {Sega},
  \citenamefont {Sbragaglia}, \citenamefont {Biferale},\ and\ \citenamefont
  {Succi}}]{sega2013regularization}%
  \BibitemOpen
  \bibfield  {author} {\bibinfo {author} {\bibfnamefont {M.}~\bibnamefont
  {Sega}}, \bibinfo {author} {\bibfnamefont {M.}~\bibnamefont {Sbragaglia}},
  \bibinfo {author} {\bibfnamefont {L.}~\bibnamefont {Biferale}}, \ and\
  \bibinfo {author} {\bibfnamefont {S.}~\bibnamefont {Succi}},\ }\bibfield
  {title} {\enquote {\bibinfo {title} {Regularization of the slip length
  divergence in water nanoflows by inhomogeneities at the angstrom scale},}\
  }\href@noop {} {\bibfield  {journal} {\bibinfo  {journal} {Soft Matter}\
  }\textbf {\bibinfo {volume} {9}},\ \bibinfo {pages} {8526--8531} (\bibinfo
  {year} {2013})}\BibitemShut {NoStop}%
\bibitem [{\citenamefont {Sega}\ \emph {et~al.}(2015)\citenamefont {Sega},
  \citenamefont {Sbragaglia}, \citenamefont {Biferale},\ and\ \citenamefont
  {Succi}}]{sega2015importance}%
  \BibitemOpen
  \bibfield  {author} {\bibinfo {author} {\bibfnamefont {M.}~\bibnamefont
  {Sega}}, \bibinfo {author} {\bibfnamefont {M.}~\bibnamefont {Sbragaglia}},
  \bibinfo {author} {\bibfnamefont {L.}~\bibnamefont {Biferale}}, \ and\
  \bibinfo {author} {\bibfnamefont {S.}~\bibnamefont {Succi}},\ }\bibfield
  {title} {\enquote {\bibinfo {title} {The importance of chemical potential in
  the determination of water slip in nanochannels},}\ }\href@noop {} {\bibfield
   {journal} {\bibinfo  {journal} {The European Physical Journal E}\ }\textbf
  {\bibinfo {volume} {38}},\ \bibinfo {pages} {1--7} (\bibinfo {year}
  {2015})}\BibitemShut {NoStop}%
\bibitem [{\citenamefont {Hilaire}\ \emph {et~al.}(2023)\citenamefont
  {Hilaire}, \citenamefont {Siboulet}, \citenamefont {Charton},\ and\
  \citenamefont {Dufr{\^e}che}}]{hilaire2023liquid}%
  \BibitemOpen
  \bibfield  {author} {\bibinfo {author} {\bibfnamefont {L.}~\bibnamefont
  {Hilaire}}, \bibinfo {author} {\bibfnamefont {B.}~\bibnamefont {Siboulet}},
  \bibinfo {author} {\bibfnamefont {S.}~\bibnamefont {Charton}}, \ and\
  \bibinfo {author} {\bibfnamefont {J.-F.}\ \bibnamefont {Dufr{\^e}che}},\
  }\bibfield  {title} {\enquote {\bibinfo {title} {Liquid--liquid flow at
  nanoscale: Slip and hydrodynamic boundary conditions},}\ }\href@noop {}
  {\bibfield  {journal} {\bibinfo  {journal} {Langmuir}\ } (\bibinfo {year}
  {2023})}\BibitemShut {NoStop}%
\bibitem [{\citenamefont {Eyring}(1936)}]{eyring_viscosity_1936}%
  \BibitemOpen
  \bibfield  {author} {\bibinfo {author} {\bibfnamefont {H.}~\bibnamefont
  {Eyring}},\ }\bibfield  {title} {\enquote {\bibinfo {title} {Viscosity,
  plasticity, and diffusion as examples of absolute reaction rates},}\ }\href
  {\doibase 10.1063/1.1749836} {\bibfield  {journal} {\bibinfo  {journal} {The
  Journal of chemical physics}\ }\textbf {\bibinfo {volume} {4}},\ \bibinfo
  {pages} {283--291} (\bibinfo {year} {1936})}\BibitemShut {NoStop}%
\bibitem [{\citenamefont {Benjamin}(1996)}]{benjamin_chemical_1996}%
  \BibitemOpen
  \bibfield  {author} {\bibinfo {author} {\bibfnamefont {I.}~\bibnamefont
  {Benjamin}},\ }\bibfield  {title} {\enquote {\bibinfo {title} {Chemical
  reactions and solvation at liquid interfaces: {{A}} microscopic
  perspective},}\ }\href {\doibase 10.1021/cr950230+} {\bibfield  {journal}
  {\bibinfo  {journal} {Chemical reviews}\ }\textbf {\bibinfo {volume} {96}},\
  \bibinfo {pages} {1449--1476} (\bibinfo {year} {1996})}\BibitemShut {NoStop}%
\bibitem [{\citenamefont
  {{Finlayson-Pitts}}(2009)}]{finlayson-pitts_reactions_2009}%
  \BibitemOpen
  \bibfield  {author} {\bibinfo {author} {\bibfnamefont {B.~J.}\ \bibnamefont
  {{Finlayson-Pitts}}},\ }\bibfield  {title} {\enquote {\bibinfo {title}
  {Reactions at surfaces in the atmosphere: Integration of experiments and
  theory as necessary (but not necessarily sufficient) for predicting the
  physical chemistry of aerosols},}\ }\href {\doibase 10.1039/B906540G}
  {\bibfield  {journal} {\bibinfo  {journal} {Physical Chemistry Chemical
  Physics}\ }\textbf {\bibinfo {volume} {11}},\ \bibinfo {pages} {7760--7779}
  (\bibinfo {year} {2009})}\BibitemShut {NoStop}%
\bibitem [{\citenamefont {Taccardi}\ \emph {et~al.}(2017)\citenamefont
  {Taccardi}, \citenamefont {Grabau}, \citenamefont {Debuschewitz},
  \citenamefont {Distaso}, \citenamefont {Brandl}, \citenamefont {Hock},
  \citenamefont {Maier}, \citenamefont {Papp}, \citenamefont {Erhard},\ and\
  \citenamefont {Neiss}}]{taccardi_gallium-rich_2017}%
  \BibitemOpen
  \bibfield  {author} {\bibinfo {author} {\bibfnamefont {N.}~\bibnamefont
  {Taccardi}}, \bibinfo {author} {\bibfnamefont {M.}~\bibnamefont {Grabau}},
  \bibinfo {author} {\bibfnamefont {J.}~\bibnamefont {Debuschewitz}}, \bibinfo
  {author} {\bibfnamefont {M.}~\bibnamefont {Distaso}}, \bibinfo {author}
  {\bibfnamefont {M.}~\bibnamefont {Brandl}}, \bibinfo {author} {\bibfnamefont
  {R.}~\bibnamefont {Hock}}, \bibinfo {author} {\bibfnamefont {F.}~\bibnamefont
  {Maier}}, \bibinfo {author} {\bibfnamefont {C.}~\bibnamefont {Papp}},
  \bibinfo {author} {\bibfnamefont {J.}~\bibnamefont {Erhard}}, \ and\ \bibinfo
  {author} {\bibfnamefont {C.}~\bibnamefont {Neiss}},\ }\bibfield  {title}
  {\enquote {\bibinfo {title} {Gallium-rich {{Pd}}\textendash{{Ga}} phases as
  supported liquid metal catalysts},}\ }\href {\doibase 10.1038/nchem.2822}
  {\bibfield  {journal} {\bibinfo  {journal} {Nature chemistry}\ }\textbf
  {\bibinfo {volume} {9}},\ \bibinfo {pages} {862--867} (\bibinfo {year}
  {2017})}\BibitemShut {NoStop}%
\end{thebibliography}%


%

\end{document}